\documentclass[prb,aps,showpacs,unsortedaddress,superscriptaddress,amsmath,amssymb,twocolumn,floatfix]{revtex4}
\usepackage{graphicx}
\usepackage{amssymb}
\usepackage{amsbsy}
\usepackage[usenames]{color}
\usepackage{amsmath}
\usepackage{amsthm}

\begin{document}

\title{Shear viscosity of strongly interacting fermionic quantum fluids}
\author{Nandan Pakhira}
\author{Ross H. McKenzie}
\email{email: r.mckenzie@uq.edu.au}
\homepage{URL: condensedconcepts.blogspot.com}
\affiliation{School of Mathematics and Physics, The University of Queensland, Brisbane, QLD 4072, Australia.}
\begin{abstract}
Eighty years ago Eyring proposed that the shear viscosity of a liquid, $\eta$, has a quantum limit $\eta \gtrsim n\hbar$ where $n$ is the density of the fluid. Using holographic 
duality and the AdS/CFT correspondence in string theory Kovtun, Son, and Starinets (KSS) conjectured a universal bound $\frac{\eta}{s}\geq \frac{\hbar}{4\pi k_{B}}$ for the ratio 
between the shear viscosity and the entropy density, $s$. Using Dynamical Mean-Field Theory (DMFT) we calculate the shear viscosity and entropy density for a fermionic fluid 
described by a single band Hubbard model at half filling. Our calculated shear viscosity as a function of temperature is compared with experimental data for liquid $^{3}$He. At 
low temperature the shear viscosity is found to be well above the quantum limit and is proportional to the characteristic Fermi liquid $1/T^{2}$ dependence, where $T$ is the 
temperature. With increasing temperature and interaction strength $U$ there is significant deviation from the Fermi liquid form. Also, the shear viscosity violates the quantum 
limit near the crossover from coherent quasi-particle based transport to incoherent transport (the bad metal regime). Finally, the ratio of the shear viscosity to the entropy 
density is found to be comparable to the  KSS bound for parameters appropriate to liquid $^{3}$He. However, this bound is found to be strongly violated in the bad metal regime
for parameters appropriate to lattice electronic systems such as organic charge transfer salts.
\end{abstract}
\maketitle

\section{Introduction}
The viscosity of a fluid is a measure of its resistance to externally applied \textit{shear} or \textit{tensile stress}. The shear viscosity of a fluid measures the resistance 
of a fluid to \textit{shear flows}, where adjacent layers of a fluid move parallel to each other but with different speeds. The differential speed between different layers will
give rise to friction between different layers which will resist their relative motion. This is known as the \textit{viscous drag}. For example, the viscous drag force per unit 
area in the $x$-direction, $\tau_{xy}$, due to velocity gradient $\partial u_{x}(y)/\partial y$ in the perpendicular $y$-direction is given by :
\begin{eqnarray}
\tau_{xy} = -\eta \frac{\partial u_{x}(y)}{\partial y},
\label{eq:Fshear}
\end{eqnarray}  
where $\eta$ is the coefficient of shear viscosity. The SI unit of shear viscosity is Pascal-seconds (Pa.s) equivalent to \textit{Newton-second per square meter} 
($\textrm{N.s m}^{-2}$). The shear viscosity of water is about $10^{-3} \textrm{(Pa.s)}$ at room temperature whereas the shear viscosity of highly viscous fluids such as  glasses 
near the glass transition temperature can be as large at 10$^{13}$ Pa.s.

For fluids $\eta$ can be measured through Stokes law for sound attenuation: \cite{Bhatia}
\begin{eqnarray}
\alpha=\frac{2\omega^{2}\eta}{3\rho c_{s}^{3}}
\label{eqn-stokes}
\end{eqnarray}  
where $\alpha$ is the rate of attenuation, $\rho$ is the mass density of the fluid, $\omega$ and $c_{s}$ are the frequency and velocity of sound in the medium, respectively. This 
equation has been used to determine the shear viscosity as a function of temperature for liquid $^3$He (a correlated neutral fermion fluid). Extensive experimental data has been 
reviewed by Huang et al.\cite{HuangCryogenics2012}.

The shear viscosity for an electron gas in metals, calculated from solution of the Boltzmann equation, is given by~\cite{SteinbergPR1958}
\begin{eqnarray}
\eta = \frac{1}{5} n\hbar k_{F}\ell,
\label{eqn-fem}
\end{eqnarray}
where $n$ is the density of electrons, $k_{F}$ is the Fermi velocity, and $\ell$ is the electronic \textit{mean free path}, respectively. In the quasi-particle regime of 
transport $k_{F}\ell \gg 1$, i.e., the mean free path is much larger than the lattice spacing, $a\sim k_{F}^{-1}$. Hence, in analogy with the Mott-Ioffe-Regel (MIR) limit, 
$\sigma_{MIR}=\frac{e^{2}}{h a}$, for minimum metallic conductivity, we can conjecture a lower limit for the shear viscosity, $\eta_{q}$ :
\begin{eqnarray}
\eta_{q}=\frac{1}{5} n\hbar
\end{eqnarray}
corresponding to the case where the electronic mean free path becomes comparable to lattice spacing. Also, a comparable limit $\eta \gtrsim n\hbar$ was proposed by 
Eyring~\cite{EyringJCP1936} almost 80 years ago. For a large class of strongly correlated systems like $3d$ transition metal oxide compounds, organic charge transfer salts such 
as $\kappa$-(BEDT-TTF)$_{2}$X, the MIR limit is violated \cite{MerinoPRB2000,HusseyPhilMag2004,Calandra} and the coherent quasi-particle based transport picture breaks down, i.e. 
$\ell < a$. Similarly, we might expect that in the incoherent regime of transport the shear viscosity, $\eta$, could violate the quantum limit to coherent transport, i.e. 
$\eta < \eta_{q}$. 

Recently a string theory based approach has been proposed to understand incoherent quantum transport in strongly correlated electron systems, especially the strange metal regime 
of doped cuprates~\cite{SachdevJPcond2009,FaulknerScience2010,LiuPT2012,HartnollNatPhys2014,Zaanen}. The key idea of this method is to map a strongly coupled conformal field 
theory (CFT) to weakly coupled gravity in the anti-de Sitter (AdS) space in higher dimension~\cite{Maldacena1998}. This is known as the holographic duality or AdS/CFT 
correspondence. Furthermore, event horizon dynamics of a black hole in the anti-de Sitter space can be mapped to the dynamics of classical fluids. Using the AdS/CFT correspondence Kovtun, Son, and Starinets (KSS)~\cite{KovtunPRL2005} calculated the ratio, $\eta/s$, of the shear viscosity ($\eta$) and the entropy density ($s$) in a specific string theory 
model (type IIB) and proposed a universal lower bound for the ratio
\begin{eqnarray} 
\frac{\eta}{s} \geq \frac{\hbar}{4\pi k_{B}}
\end{eqnarray}
{\it in any material or field theory}. This bound is found to be well respected in classical fluids like water and quantum fluids like the quark-gluon plasma created in the 
Relativistic Heavy Ion Collider (RHIC)~\cite{Shuryak2004}, ultracold atomic Fermi gases in the unitary limit of scattering~\cite{CaoScience2011}, and by theoretical calculations 
for graphene~\cite{MullerPRL2009} and for ultracold atomic Fermi gases~\cite{Chafin,Bulgac}. 
It has recently been found that for the strongly interacting Fermi gas, both experimentally~\cite{ElliottPRL2014} and theoretically \cite{Wlazlowski}, that the viscosity-entropy 
ratio is a minimum, not at unitarity, but on the BEC side, with a minimum value of $\simeq 0.2 \hbar/k_B$. Possible violations of the KSS bound have been discussed for higher 
derivative versions of gravity and with the inclusion of massive quarks~\cite{Myers}. 
 
In a recent calculation we tested a related but distinct bound on charge diffusivity, $D \geq \frac{\hbar v_{F}^{2}}{k_{B}T}$, where $v_{F}$ is the Fermi velocity, proposed by 
Hartnoll~\cite{HartnollNatPhys2014}. We found~\cite{PakhiraPRB2015} clear violation of this bound in the strong coupling (bad metal) regime of the Hubbard model. In the present 
paper we calculate the shear viscosity in a single band Hubbard model and explore possible violations of the conjectured quantum bounds on $\eta$ and $\eta/s$.

Overall we find that the scale of the viscosity in a correlated band system with a lattice constant $a$ in $d$ dimensions is set by
\begin{eqnarray} 
\eta_{0}^{b} \equiv \frac{\hbar}{a^{d}} \left(\frac{m}{m_{b}}\right)^{2}\left[\frac{\pi}{2d(d-1)}\right]
\label{eqn-etab0}
\end{eqnarray}  
where $m$ is the free fermion mass and $m_{b} \equiv \hbar^{2}/(a^{2} E_{b})$ is a mass scale determined by some energy scale $E_{b}$ defined by the band structure, such as the 
half-band width $W$ or the rescaled hopping integral $t^*$ for a hypercubic lattice. A detailed derivation of the expression in Eq.(\ref{eqn-etab0}) will be provided in the 
following sections. We show that for a lattice system $\eta_0^b$ is the relevant scale for the analogue of the Mott-Ioffe-Regel limit. We will see that the presence of this new 
scale, absent in a conformally invariant system, can increase the likelihood of violation of the KSS bound.

The organization of the paper is as follows. In  Section II we introduce the Kubo formula for calculation of the shear viscosity using linear response theory. In Section III we 
briefly describe the Dynamical Mean-Field Theory (DMFT) approach for calculating properties of a single band Hubbard model and the iterated perturbation theory (IPT) based 
approach used to treat the DMFT self-consistency for the associated single impurity Anderson model. In the same section we introduce calculation of the shear viscosity and entropy 
density in  DMFT. In Section IV we first briefly review experimental results for the temperature dependence of the shear viscosity of liquid $^3$He and its possible description by 
a Hubbard model. In Section V we  show our results for the Hubbard model on the Bethe and hypercubic lattices at half filling. Similar results are obtained for both lattices. We 
compare our calculations to experimental results for liquid $^3$He. The temperature dependence of the ratio of the viscosity to the entropy density is calculated. It is found that 
in the bad metal regime near the Mott metal-insulator transition this ratio can be smaller than the KSS bound. In Section VI we discuss about experimental measurement of shear 
viscosity in charged systems and finally we conclude in Section VII.  

\section{Shear Viscosity}
Nonrelativistic simple fluids are characterized by the conserved mass density $\rho$, the momentum density $\boldsymbol{\pi}$ and the energy density $\mathcal{E}$. These quantities will satisfy following conservation laws~\cite{SchaferAnnRevNuclPart2014} :
\begin{eqnarray}
\label{eq:drhodt}
\frac{\partial \rho}{\partial t} &=& -\boldsymbol{\nabla.\pi}\\
\label{eq:dpidt}
\frac{\partial \pi_{i}}{\partial t} &=& -\nabla_{j}\Pi_{ij}\\
\label{eq:dEdt}
\frac{\partial \mathcal{E}}{\partial t} &=& -\boldsymbol{\nabla.j}^{\epsilon}
\end{eqnarray} 
where $\Pi_{ij}$ is the momentum current density that the following discussion shows is central to the shear viscosity. As a consequence, in analogy with the case of Ohm's law for 
electrical conductivity : $j^{e}_{\alpha}=\sigma_{\alpha\beta}E_{\beta}\equiv -\sigma_{\alpha\beta}\partial_{\beta}\phi(\mathbf{r})$, the generalized Newton's law for shear flow is
\begin{eqnarray}
\Pi_{\alpha\beta}=-\eta_{\alpha\beta\gamma\delta}(T) \partial_{\gamma}u_{\delta}(\mathbf{r}),
\end{eqnarray}
where $\eta_{\alpha\beta\gamma\delta}$ is a viscosity tensor. In particular, the momentum current density $\Pi_{xy}$ in the presence of a transverse velocity gradient 
$\frac{\partial u_{x}(y)}{\partial y}$ is given by~\cite{BruunPRA2007} 
\begin{eqnarray}
\Pi_{xy}=-\eta \frac{\partial u_{x}}{\partial y},
\end{eqnarray}
where $\eta\equiv\eta_{xyxy}$ is the coefficient of shear viscosity for an isotropic fluid.
 
The velocity field $u_{x}(y)$ gives rise to a perturbation with Hamiltonian
\begin{eqnarray}
H' = -\int d^{3}r \ u_{x}(y) \hat{\pi}_{x}(\mathbf{r}) = \frac{1}{i\omega}\frac{\partial u_{x}}{\partial y}\int d^{3}r \hat{\Pi}_{xy}.
\label{eq:Hint}
\end{eqnarray}
To derive Eq. (\ref{eq:Hint}) we have used the conservation law in Eq.~\ref{eq:dpidt}, integration by parts, and $\hat{\boldsymbol{\pi}}(\mathbf{r},t)=\exp(-i\omega t)
\hat{\boldsymbol{\pi}}(\mathbf{r})$. The momentum current density $\Pi_{xy}$ induced by the perturbation $\hat{H}'$ can be calculated from linear response theory. The shear 
viscosity is then obtained by taking the limit $\omega\rightarrow 0$:
\begin{eqnarray}
\eta = -\lim_{\omega\rightarrow 0} \textrm{Im}\; {\Xi (\omega) \over \omega}
\label{eqn:eta0}
\end{eqnarray} 
with~\cite{SchaferAnnRevNuclPart2014,TaylorPRA2010} 
\begin{eqnarray}
\Xi(\omega) = -i\frac{\hbar}{\nu}\int d^{3}r \ dt \ e^{i\omega t} \ \theta(t) \langle \left[\hat{\Pi}_{xy}(\mathbf{r},t),\hat{\Pi}_{xy}(0,0)\right] \rangle
\end{eqnarray}
where $\nu = a^3$ is the unit cell volume and $\theta(t)$ is the Heaviside step function. This formula is the analogue of the Kubo expression for the electrical conductivity 
involving the current-current correlation function.

For a Fermi gas with a quadratic energy dispersion the momentum current density operator is given by~\cite{BruunPRA2007}
\begin{eqnarray}
\hat{\Pi}_{xy} = \frac{1}{(2\pi)^{3}}\int d^{3}p \;
\frac{\hat{p}_{x}\hat{p}_{y}}{m} \;\delta f
\label{pixy}
\end{eqnarray}
where $\delta f \equiv f_{\mathbf{k}}-f_{\mathbf{k}}^{0}$ is the deviation of the distribution function from local equilibrium. For Bloch electrons in a crystal lattice
~\cite{ashcroft1976solid} in a Bloch state, $\psi_{\mathbf{k}n}(\mathbf{r})=\langle \mathbf{r} | n, \mathbf{k}\rangle$, with energy $\epsilon_{n}(\mathbf{k})$
\begin{eqnarray}
\frac{\partial \epsilon_{n}(\mathbf{k})}{\partial k_{\alpha}} = \frac{\hbar}{im}\int d\mathbf{r}\;\psi_{\mathbf{k}n}^{*}(\mathbf{r})
\frac{\partial \psi_{\mathbf{k}n}(\mathbf{r})}
{\partial r_{\alpha}}
=\frac{1}{m} \langle n,\mathbf{k} | \hat{p_\alpha} |n,\mathbf{k} \rangle 
\end{eqnarray}
where $\epsilon_{n}(\mathbf{k})$ is the energy dispersion of the $n$-th energy band. Then (\ref{pixy}) can be written 
\begin{eqnarray}
\Pi_{xy} =  \frac{m}{(2\pi)^{3}}\int d\mathbf{k}\; v_{k_{x}}v_{k_{y}}\delta f
\end{eqnarray}
with $v_{k_{\alpha}}=\frac{1}{\hbar}\frac{\partial \epsilon_{\mathbf{k}}}{\partial k_{\alpha}}$ being the velocity of the Bloch electron. Using \textit{deformation potential} 
theory~\cite{KhanPRB1984} a similar result was found by Khan and Allen~\cite{KhanPRB1987} when investigating sound attenuation by electrons in metals.

It should be pointed out that in a general fluid  there are two terms in the stress energy tensor: one associated with the kinetic energy and the second with the interparticle 
interaction. In dense classical liquids the terms in the Kubo formula due to the interaction term dominates and is associated with Einstein-Stokes relation where the viscosity 
is inversely proportional to the particle self-diffusion constant. In contrast, in dilute gases and fluids the kinetic term dominates and the shear viscosity scales with the 
diffusion constant and scattering time~\cite{Rah}. However, for a zero-range interaction, as in the unitary Fermi gas (and presumably in the Hubbard model), it can be shown that 
the potential term in the stress tensor does not contribute to the shear viscosity.For a discussion of the above see around Eq. 7 in Ref.~\onlinecite{Enss}.
\section{Dynamical mean field theory}
We consider the single band Hubbard model with nearest neighbor hopping, described by the Hamiltonian
\begin{eqnarray}
H = -t\sum_{\langle ij \rangle,\sigma}(c_{i\sigma}^{\dagger}c_{j\sigma}+\textrm{H.c.})-\mu\sum_{i,\sigma}n_{i\sigma}+U\sum_{i}n_{i\uparrow}n_{i\downarrow},
\end{eqnarray}
where $n_{i\sigma}=c_{i\sigma}^{\dagger}c_{i\sigma}$, $t$ is the hopping amplitude, $\mu$ is the chemical potential, and $U$ is the Coulomb repulsion when a given site is doubly 
occupied by two fermions with opposite spin configuration. Despite its simplicity this model has no exact solution except in one dimension. The study of this model in higher 
dimension involves various approximations. However, as in the case of classical mean field theory for the nearest neighbour Ising model, in the limit of large dimension, 
$d\rightarrow\infty$ the model reduces to an effective single site model provided we do the scaling $t\rightarrow t^{*}/\sqrt{2d}$ on a $d$-dimensional hypercubic 
lattice~\cite{MetznerPRL1989}. Under this approximation we neglect all spatial fluctuations yet fully retain local quantum dynamics. The self-energy $\Sigma_{ij}(\omega)$ for the 
lattice model then becomes \textit{local}, i.e. $\Sigma_{ij}(\omega)=\Sigma(\omega)\delta_{ij}$. This is known as the Dynamical Mean-Field Theory~\cite{GeorgesRMP1995} (DMFT) 
approximation. 

It has been found that DMFT gives a good description of the correlation driven Mott metal-insulator transition observed in $3d$ transition metal oxides and the crossover from a 
coherent Fermi liquid to incoherent bad metal state with increasing temperature~\cite{MerinoPRB2000}. Furthermore, DMFT has also been found to provide quantitative description of 
the resistivity~\cite{LimelettePRL2003} and the frequency dependent optical conductivity~\cite{MerinoPRL2008} for organic charge-transfer salts that can be described by a half-
filled two-dimensional Hubbard model on an anisotropic triangular lattice~\cite{PowellRepProgPhys2011}. DMFT combined with electronic structure calculations based on density 
functional theory (DFT) has given an excellent description of a large class of transition metal and rare earth compounds~\cite{KotliarRMP2006}.     

The lattice problem under DMFT can be mapped onto an effective single impurity Anderson model~\cite{GeorgesRMP1995} :
\begin{eqnarray}
H_{\textrm{imp}}&=&\sum_{l,\sigma}(\tilde{\epsilon}_{l}-\mu)c_{l\sigma}^{\dagger}c_{l\sigma}+\sum_{l,\sigma}(V_{l}c_{l\sigma}^{\dagger}d_{0\sigma}+\textrm{H.c.})\nonumber \\
                & & -\mu\sum_{\sigma}n_{d0\sigma}+U n_{d0\uparrow}n_{d0\downarrow},
\end{eqnarray}
where $n_{d0\sigma}=d_{0\sigma}^{\dagger}d_{0\sigma}$. The operators $d_{0\sigma}^{\dagger}$ and $d_{0\sigma}$ characterizes a local site and $\{c_{l\sigma}^{\dagger},c_{l\sigma}\}$ characterizes the effective bath arising from fermions at all other sites. It is important to mention that the fictitious bath dispersion $\tilde{\epsilon}_{l}$ has no relation to 
the lattice dispersion, $\epsilon_{\mathbf{k}}$.

The solution of the impurity problem is the toughest part and usually involves use of numerical methods such as Quantum Monte Carlo (QMC), exact diagonalization (ED), or the
numerical renormalization group (NRG). We use iterated perturbation theory (IPT)~\cite{ZhangPRL1993,KajueterPRL1996} as it is semi-analytical, easy to implement, computationally 
cheap and fast. Yet IPT captures the essential physics in the parameter regime $U < 0.8U_{c}$, where $U_{c}$ is the critical value of $U$ at which the zero temperature Mott 
metal-insulator transition happens. Except in close proximity of the Mott transition IPT was found to be in good agreement with results from other impurity solvers such as the 
numerical renormalization group (NRG)~\cite{BullaPRL1999} and continuous time quantum Monte Carlo (CTQMC)~\cite{TerletskaPRL2011}. In the next sub-section we discuss DMFT self-
consistency using IPT.      

We briefly mention why it is appropriate to compare the results of AdS/CFT to a calculation involving DMFT. The latter becomes exact in infinite dimension. Generally, AdS/CFT is 
concerned with finite-dimensional quantum field theories. However, it is found that in certain parameter regimes DMFT can accurately give a quantitative description of 
quasi-two-dimensional metals near the Mott insulator. Furthermore, connections have been made between the results of AdS/CFT and the infinite-dimensional limit of a model for a 
gapless spin liquid~\cite{SachdevPRL}. We know that the DMFT approximation reduces a lattice problem to an effective local impurity problem which captures local correlation 
effects. On the other hand the AdS/CFT correspondence maps a strongly coupled field theory to a problem of fluid mechanics and fluids are characterised by short range 
correlations. So we might expect DMFT based description of quantum transport of lattice electronic systems will be closely related to quantum transport in the hydrodynamic regime 
of a strongly coupled field theory. 
\subsection{Iterated perturbation theory}
The irreducible self-energy in IPT is approximated using the second order (in $U$) polarization bubble involving fully interacting bath Green's function $G_{0}(\omega)$. The 
self-energy under this approximation can be shown (using moment expansion of the interacting density of states) to smoothly interpolate between the atomic limit $t=0$ and the 
weak-coupling limit $U\rightarrow 0$. In the following paragraph we briefly discuss DMFT self-consistency  using IPT as the impurity solver. As we are interested in calculating 
transport properties we work with real frequencies, as against the imaginary frequency formulation that requires analytical continuation of imaginary frequency data to real 
frequency. 

(i) For a given lattice density of states $N_{0}(\epsilon)$ and self-energy $\Sigma(\omega)$ the \textit{local} Green's function is given by
\begin{eqnarray}
G(\omega) = \int_{-\infty}^{+\infty}\frac{N_{0}(\epsilon)d\epsilon}{\omega^{+}+\mu-\epsilon-\Sigma(\omega^{+})},
\end{eqnarray}
where $\mu$ is the local chemical potential and $\omega^{+}=\omega+i\delta$ with $\delta > 0$.\\
(ii) From the knowledge of the local Green's function, $G(\omega)$, we can calculate the \textit{bath hybridization} function, $\Delta(\omega)$ by using
\begin{eqnarray}
\Delta(\omega)=\omega^{+}+\mu-\Sigma(\omega)-G^{-1}(\omega)
\label{eq:defnHybridzn}
\end{eqnarray} 
(iii) Subsequently using \textit{bath hybridization} we can calculate \textit{bath} Green's function as
\begin{eqnarray}
G_{0}(\omega)=\frac{1}{\omega+\mu_{0}-\Delta(\omega)}.
\label{eq:defnBathGreenFn}
\end{eqnarray}
The  parameter $\mu_{0}=\mu-Un$ is the bath chemical potential and it vanishes at half-filling for the particle-hole symmetric case, which we consider in the present study.\\
(iv) The new self-energy can be calculated using IPT ansatz~\cite{KajueterPRL1996} as
\begin{eqnarray}
\Sigma(\omega)=Un+\frac{A\Sigma^{(2)}(\omega)}{1-B\Sigma^{(2)}(\omega)}
\end{eqnarray} 
where
\begin{eqnarray}
A=\frac{n(1-n)}{n_{0}(1-n_{0})}\;;\hspace*{0.5cm} B=\frac{U(1-n)-\mu+\mu_{0}}{n_{0}(1-n_{0})U^{2}} 
\end{eqnarray}
and $n$, $n_{0}$ are the local and bath particle numbers, respectively. $\Sigma^{(2)}(\omega)$ is the self energy from second order
perturbation theory and is given by
\begin{eqnarray}
\Sigma^{(2)}(\omega) &=& U^{2}\!\!\int\limits_{-\infty}^{+\infty}\prod_{i=1}^{3}\left(d\epsilon_{i}\rho_{0}(\epsilon_{i})\right)
\left[\frac{n_{F}(-\epsilon_{1})n_{F}(\epsilon_{2})n_{F}(-\epsilon_{3})}{\omega+i\delta-\epsilon_{1}+\epsilon_{2}-\epsilon_{3}}\right . \nonumber \\
&&\hspace*{1cm}\left . +\frac{n_{F}(\epsilon_{1})n_{F}(-\epsilon_{2})n_{F}(\epsilon_{3})}{\omega+i\delta-\epsilon_{1}+\epsilon_{2}-\epsilon_{3}}\right],
\end{eqnarray}
where $\rho_{0}(\omega)=-\frac{1}{\pi}\textrm{Im}[G_{0}(\omega^{+})]$ and $\delta\rightarrow 0^{+}$. We iterate (i)-(iv) until the desired self-consistency in self-energy and 
other physical quantities are achieved. Here we consider the particle-hole symmetric case at half filling $n=1$. In this case $\mu=\frac{U}{2}$ for all $U$ and $T$.
\subsection{Shear viscosity in DMFT}
Using the self-consistent self-energy we can calculate the shear viscosity. In the limit of $d\rightarrow\infty$ all vertex corrections to two-body correlation functions drops 
out~\cite{KhuranaPRL1990} and the temperature dependent coefficient of shear viscosity, $\eta(T)$, given by the Kubo formula Eq. (\ref{eqn:eta0}) can be calculated using a simple 
polarization bubble as
\begin{eqnarray}
\eta(T)=\frac{\pi\hbar}{\nu}\int_{-\infty}^{+\infty}d\omega\left[-\frac{\partial n_{F}(\omega)}{\partial\omega}\right]
       \int_{-\infty}^{+\infty} d\epsilon \;\Theta_{xy}(\epsilon) A^{2}(\omega,\epsilon),
\label{eqn:etadmft}
\end{eqnarray} 
where $\nu=a^{d}$ is the volume of the unit cell of a $d$-dimensional hypercubic lattice with lattice constant $a$,
\begin{eqnarray}
\label{eq:defnSpctrlDen}
A(\omega,\epsilon) &=& -\frac{1}{\pi}\textrm{Im}\left[\frac{1}{\omega^{+}+\mu-\epsilon-\Sigma(\omega^{+})}\right]\\
\label{eq:FermiFn}
     n_{F}(\omega) &=& \frac{1}{e^{\beta\omega}+1}
\end{eqnarray}
are the spectral density and Fermi function, respectively.
\begin{eqnarray}
\Theta_{xy}(\epsilon) = \frac{m^2}{N} \sum_{\mathbf{k}} v_{kx}^{2}v_{ky}^{2}\delta(\epsilon-\epsilon_{\mathbf{k}})
\label{eqn:thetaxy}
\end{eqnarray}
with $v_{k\alpha}=\frac{1}{\hbar}\frac{\partial \epsilon_{\mathbf{k}}}{\partial k_{\alpha}}$ is the {\it transport density of states for the shear viscosity} and $N$ is the number 
of lattice sites. Following a similar procedure to that in Ref.~\onlinecite{ArsenaultPRB2013} we can show that the transport density of states for shear viscosity for a $d$-
dimensional hypercubic lattice with nearest neighbour hopping is given by
\begin{eqnarray}
\Theta_{xy}(\epsilon) &=& \frac{\gamma^{2}}{2d(d-1)}\left[-\frac{3}{2} M_{3}(\epsilon) + 2 \epsilon M_{2}(\epsilon)\right . \nonumber\\
                      &+& \left . (4 d t^{2} -\frac{1}{2} \epsilon^{2}) M_{1}(\epsilon)-4 t^2 \epsilon M_{0}(\epsilon)\right]
\label{eqn:thetaxy2}
\end{eqnarray}
where $\gamma=\frac{ma^{2}}{\hbar^{2}}$ and
\begin{equation}
M_{n}(\epsilon) \equiv \int\limits_{-\infty}^{\epsilon} z^n N_{0}(z) dz
\label{eqn:dos-int}
\end{equation}
where $N_{0}(\epsilon)=\sum_{\mathbf{k}}\delta(\epsilon-\epsilon_{\mathbf{k}})$ is the \textit{density of states} per spin. In the Appendix we give a detailed derivation of this 
important result. In the following sub-sections we explicitly evaluate this expression for the hypercubic lattice  and Bethe lattice cases.  

One should consider how the vertex corrections could modify the DMFT results in finite dimensions. For the unitary fermi gas vertex corrections increase the viscosity by a factor 
of about 2.6 [compare the discussion below equation (54) in Reference~\onlinecite{Enss}]. For the quark-gluon plasmon in the theory of Quantum Chromodynamics (QCD) at high 
temperatures vertex corrections significantly increase the viscosity, changing the functional dependence on the coupling constant [compare equations (4.25) and (4.26) in 
Reference \onlinecite{Aarts}]. In a two-dimensional Fermi liquid the vertex corrections have been shown~\cite{Novikov} to be of the order of $(\ln(E_F/T))^3$. For a doped Hubbard 
model it was found in a Dynamical Cluster Approximation calculation based on a four site cluster that the vertex corrections to the optical conductivity were not significant, 
except very close to the Mott insulator~\cite{Lin}. A study of the same model using a two-particle self-consistent approach found that vertex corrections changed the calculated 
resistivity by less than a factor of two~\cite{BergeronPRB2011}. In light of the above it seems unlikely that vertex corrections would increase the viscosity by more than an 
order of magnitude compared to the DMFT results.
\subsubsection{Hypercubic lattice case}
For hypercubic lattice in the limit of $d\rightarrow\infty$ we have $N_{0}(\epsilon)=\frac{1}{\sqrt{\pi}t^{*}}\exp[-\epsilon^{2}/t^{*2}]$ for the density of states. It is 
important to mention that the chosen density of states in the limit $d\rightarrow\infty$ requires the scaling : $2t\rightarrow t^{*}/\sqrt{d}$. The transport density of states 
for the shear viscosity is then given by 
\begin{eqnarray}
\Theta_{xy}(\epsilon)&=&\frac{\gamma^{2}t^{*3}}{2d(d-1)}\left[-3I_{3}(\tilde{\epsilon})+4\tilde{\epsilon}I_{2}(\tilde{\epsilon})
                       -\tilde{\epsilon}^{2}I_{1}(\tilde{\epsilon})\right .\nonumber \\
                     & &\hspace*{2.9cm}\left . +2I_{1}(\tilde{\epsilon})-2\tilde{\epsilon}I_{0}(\tilde{\epsilon})\right]
\label{eq:TrDOSright}
\end{eqnarray}
where
\begin{eqnarray}
I_{n}(x)=\frac{1}{\sqrt{\pi}}\int_{-\infty}^{x} u^{n} e^{-u^{2}} du
\end{eqnarray}
are the dimensionless integrals and $\tilde{\epsilon}=\epsilon/t^{*}$ is the dimensionless energy. 

We define the scaled dimensionless transport density of states for the shear viscosity, 
$\bar{\Theta}(\tilde{\epsilon})$, as :
\begin{eqnarray}
\bar{\Theta}_{xy}(\tilde{\epsilon})=-3I_{3}(\tilde{\epsilon})+4\tilde{\epsilon}I_{2}(\tilde{\epsilon})-(\tilde{\epsilon}^{2}-2)I_{1}(\tilde{\epsilon})
                                 -2\tilde{\epsilon}I_{0}(\tilde{\epsilon}).
\end{eqnarray}
Using the exact integrals
\begin{eqnarray}
& &\int_{-\infty}^{\epsilon} (z^{2}-\frac{1}{2})e^{-z^{2}} dz = -\frac{1}{2}\epsilon e^{-\epsilon^{2}}\\
& &\int_{-\infty}^{\epsilon} z e^{-z^{2}} dz = -\frac{1}{2} e^{-\epsilon^{2}}\\
& &\int_{-\infty}^{\epsilon} z^{3} e^{-z^{2}} dz = -\frac{1}{2}(\epsilon^{2}+1)e^{-\epsilon^{2}}
\end{eqnarray}
we get
\begin{eqnarray}
\bar{\Theta}_{xy}(\tilde{\epsilon})=\frac{1}{2}N_{0}(\tilde{\epsilon}) = \frac{1}{2\sqrt{\pi}}e^{-\tilde{\epsilon}^{2}}.
\label{eq:ThetaHyp}
\end{eqnarray}
In Fig.~\ref{fig:EtaDOS} (a) we show transport density of states for viscosity for hypercubic lattice. Interestingly, the transport density of 
states for electrical conductivity for hypercubic lattice also follows a relation similar to Eq.~(\ref{eq:ThetaHyp}), as shown in Ref.~\onlinecite{ArsenaultPRB2013}.

The shear viscosity is then given by
\begin{eqnarray}
\eta = \eta_{0}^{b}\int_{-\infty}^{+\infty} d\tilde{\omega}\left[-\frac{\partial n_{F}(\tilde{\omega})}{\partial \tilde{\omega}}\right]
       \int_{-\infty}^{+\infty} d\tilde{\epsilon} \; \bar{\Theta}_{xy}(\tilde{\epsilon}) A^{2}(\tilde{\omega},\tilde{\epsilon})
\label{eqn-etadimless}
\end{eqnarray}
where the dimension full prefactor $\eta_{0}^{b}$ is given by Eq. (\ref{eqn-etab0}) with $m_b=\frac{\hbar^{2}}{a^{2}t^{*}}$ and $\tilde{\omega}=\omega/t^{*}$ is the dimensionless 
energy.     
\subsubsection{Bethe lattice case}
We consider the Bethe lattice (Cayley tree) with coordination number $z$. In the limit of infinite coordination number ($z\rightarrow\infty$), the density of states has 
\textit{semicircular} form~\cite{Economou2010Green} :
\begin{eqnarray}
 N_{0}(\epsilon)=\frac{2}{\pi W^{2}}\sqrt{W^{2}-\epsilon^{2}}\;\theta(W-|\epsilon|), 
\end{eqnarray}
where $\theta(x)$ is the Heaviside step function, $W=2t^{*}$ is the half bandwidth and the nearest neighbour hopping amplitude ($t$) in this case is scaled as 
$t\rightarrow t^{*}/\sqrt{z}$. For a Bethe lattice with coordination number $z$ the connectivity $K=z-1$ while that for a $d$-dimensional hypercubic lattice is $2d$. So, in the 
limit of large coordination number we can always do the mapping $z\leftrightarrow 2d$. Because of its tree like structure the Bethe lattice has no closed loop and hence no 
energy dispersion with Bloch wavevector {\bf k}. However,  by invoking the $f$-sum rule we can still calculate $\Theta_{xy}(\epsilon)$. For the given density of states we then 
have the following exact integrals~\cite{ArsenaultPRB2013} 
\begin{eqnarray}
M_{0}(\epsilon) &=& \frac{1}{2}\epsilon N_{0}(\epsilon)+\frac{1}{2}
                   +\frac{1}{\pi}\tan^{-1}\left[\frac{\epsilon}{\sqrt{W^{2}-\epsilon^{2}}}\right]\\
M_{1}(\epsilon) &=& -\frac{1}{3}(W^{2}-\epsilon^{2}) N_{0}(\epsilon)\\
M_{2}(\epsilon) &=& -\frac{\epsilon(W^{2}-2\epsilon^{2})}{8} N_{0}(\epsilon)+\frac{W^{2}}{8}\nonumber \\
                &+& \frac{W^{2}}{4\pi}\tan^{-1}\left[\frac{\epsilon}{\sqrt{W^{2}-\epsilon^{2}}}\right]\\
M_{3}(\epsilon) &=& \frac{1}{120} N_{0}(\epsilon)\left[18\epsilon^{4}-5\epsilon^{2}W^{2}-16W^{4}\right] 
\end{eqnarray}
Then by replacing these exact analytical integrals into the expression in Eq. (\ref{eqn:thetaxy2}) for $\Theta_{xy}(\epsilon)$ and using $W=2t\sqrt{2d}$ we get
\begin{eqnarray}
\Theta_{xy}(\epsilon) = \frac{\gamma^{2}}{240d(d-1)} N_{0}(\epsilon)\left[8W^{4}-25\epsilon^{2}W^{2}+26\epsilon^{4}\right].
\end{eqnarray}
It is interesting to mention that the constant term as well as the $\tan^{-1}\left[\frac{\epsilon}{\sqrt{W^{2}-\epsilon^{2}}}\right]$ term cancels out in the final expression 
for $\Theta_{xy}(\epsilon)$. 

\begin{figure}[!htbp]
\begin{center}
\includegraphics[scale=0.32,clip=]{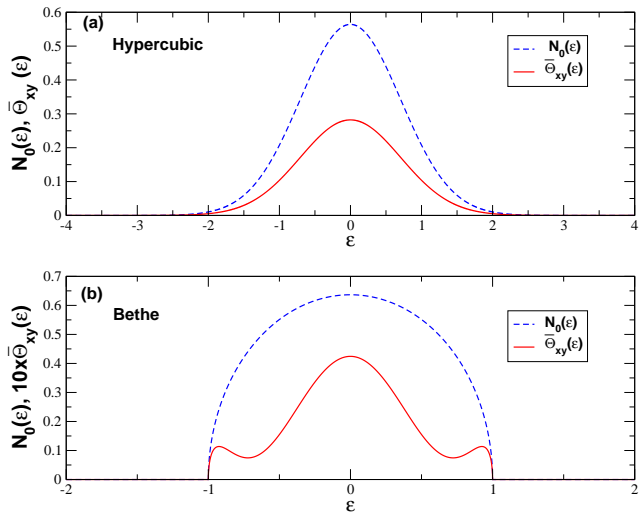}
\caption{(Color online) Density of states, $N_{0}(\epsilon)$, and scaled dimensionless transport density of states for shear viscosity, $\bar{\Theta}_{xy}(\epsilon)$, for the 
hypercubic lattice [panel (a)] and for the Bethe lattice [panel (b)]. $\bar{\Theta}_{xy}(\epsilon)$ for the Bethe lattice shows additional structures near the band edges but for 
the hypercubic lattice it is just proportional to the density of states. $\bar{\Theta}_{xy}(\epsilon)$ for the Bethe lattice has been multiplied by a factor of 10 to show both
curves in the same panels. Energies have been scaled by the effective hopping amplitude, $t^{*}$, in the case of the hypercubic lattice and by the half-band width, $W$, in the 
case of the Bethe lattice.}
\label{fig:EtaDOS}
\end{center}
\end{figure}

In Fig.~\ref{fig:EtaDOS} (b) we show the scaled dimensionless transport density of states, $\bar{\Theta}_{xy}(\tilde{\epsilon})$ :
\begin{eqnarray}
\bar{\Theta}_{xy}(\tilde{\epsilon})&=&\frac{1}{120}N_{0}(\tilde{\epsilon})\left(8-25\tilde{\epsilon}^{2}+26\tilde{\epsilon}^{4}\right)
\end{eqnarray}
with $\tilde{\epsilon}=\epsilon/W$ for the Bethe lattice. Near the band edges ($\tilde{\epsilon}=\pm 1$) $\bar{\Theta}_{xy}(\tilde{\epsilon})$ shows non-monotonic structures in 
contrast to the  density of states, $N_{0}(\tilde{\epsilon})=\frac{2}{\pi}\sqrt{1-\tilde{\epsilon}^{2}}$, which is always monotonic near the band edges.

The expression in Eq. (\ref{eqn:etadmft}) for the shear viscosity for the Bethe lattice is then given by Eq. (\ref{eqn-etadimless}) where the dimensionfull prefactor 
$\eta_{0}^{b}$ is given by Eq. (\ref{eqn-etab0}) with $m_b=\frac{\hbar^{2}}{a^{2}W}$ and $\tilde{\omega}=\omega/W$ is the dimensionless energy.    
\subsection{Entropy density}
The total internal energy in DMFT is given by\cite{Georges1993}
\begin{eqnarray}
\frac{E(T)}{N} = k_{B}T \sum_{n,\sigma} \int\limits_{-\infty}^{+\infty} d\epsilon \frac{\epsilon N_{0}(\epsilon)}{i\omega_{n}+\mu-\Sigma_{\sigma}(i\omega_{n})-\epsilon}\nonumber \\
+\frac{1}{2}\sum_{n,\sigma} \Sigma_{\sigma}(i\omega_{n})G_{\sigma}(i\omega_{n})
\label{Eq:EtotGeorges}
\end{eqnarray}
where $N$ is the total number of particles in the system, $\omega_{n}=(2n+1)\pi k_{B} T$ is the Matsubara frequency, and $N_{0}(\epsilon)$ is the non-interacting density of 
states. In the paramagnetic state Eq.~(\ref{Eq:EtotGeorges}) can be expressed as a real frequency integral
\begin{eqnarray}
\frac{E(T)}{N} &=& 2\int_{-\infty}^{+\infty} d\omega n_{F}(\omega) (\omega+\mu)A(\omega)\nonumber \\
               &+& \frac{1}{\pi}\int_{-\infty}^{+\infty} d\omega n_{F}(\omega)\mathrm{Im}\left[\Sigma(\omega^{+})G(\omega^{+})\right],
\end{eqnarray} 
where $A(\omega)=-\frac{1}{\pi}\mathrm{Im}\left[G(\omega^{+})\right]$ is the spectral function.

From $E(T)$ we can calculate the specific heat using $C_{v}(T)=\left(\frac{\partial E(T)}{\partial T}\right)_{v}$ and then we can calculate the local entropy density, $s(T)$, as
\begin{eqnarray}
s(T) = \frac{1}{\nu}\int_{0}^{T}\frac{C_{v}(T')}{T'} dT',
\end{eqnarray}
where $\nu$ is the volume of the system. The temperature dependence of the specific heat and the entropy for the half-filled Hubbard model have both been calculated previously 
using a range of impurity solvers including IPT and Quantum Monte Carlo~\cite{Georges1993,PruschkePRB1993,MoellerPRB1999}. It is found that in the metallic phase the entropy 
density is linear in temperature below the Fermi liquid coherence temperature, $T_{coh}$ and becomes of order $n k_{B}\ln(2)$ for $T \sim T_{coh}$, where $n$ is number density 
of fermions.
\subsection{Quantum limits}
The quantum limit of the 
shear viscosity, $\eta_{q}=\frac{1}{5}n\hbar$, is based on the free particle dispersion $E_{\mathbf{k}}=\frac{\hbar^{2}k^{2}}{2m}$ in the continuum limit. For a discrete lattice 
model we need to derive an appropriate quantum limit for shear viscosity. 

For temperatures and frequencies much less than the coherence scale (i.e. $T \ll T_{\mathrm{coh}}$, $\omega \ll k_{B}T_{coh}$ where $T_{coh}$ is the coherence 
temperature which is of the order of the Kondo temperature for the corresponding single impurity Anderson model) the self energy, $\Sigma(\omega)$, has the Fermi liquid form :
\begin{eqnarray}
\Sigma(\omega,T) = \omega\left(1-\frac{1}{Z}\right)-i C\left[\omega^{2}+(\pi k_{B}T)^{2}\right],
\end{eqnarray}
where $Z$ is the quasi-particle renormalization factor and $C$ is a positive constant. 

Following the procedure  in Ref.~\onlinecite{MerinoPRB2000} used to estimate the Mott-Ioffe-Regel limit for the conductivity we can show that at low temperature ($ T \ll 
T_{\mathrm{coh}}$) the shear viscosity for the hypercubic lattice is given by
\begin{eqnarray}
\eta(T) = \eta_{0}^{b}\bar{\Theta}_{xy}(0)I_{01} \pi^2 t^* \tau(T) / \hbar
\end{eqnarray}  
where $I_{01} \simeq 0.08$ is a dimensionless integral and $\hbar / \tau(T) = - {\rm Im} (\Sigma(\omega=0,T))$ is the quasi-particle decay rate. The quantum limit to shear 
viscosity will then correspond to $t^* \sim \hbar /\tau(T) $ and we will have the quantum limit to shear viscosity
\begin{eqnarray}
\eta_{q}^{lat}=\frac{1}{2{\sqrt{\pi}}} \eta_{0}^{b},
\label{Eq:EtaLatQuantumHyperCubic}
\end{eqnarray}
for the hypercubic lattice and
\begin{eqnarray}
\eta_{q}^{lat} = \frac{2}{15\pi} \eta_{0}^{b},
\label{Eq:EtaLatQuantumBethe}
\end{eqnarray}
for the Bethe lattice.
\section{Parameters for liquid $^{3}\mathrm{He}$}
We consider liquid $^{3}\mathrm{He}$ because of the availability of extensive experimental data for the temperature and pressure dependence of the shear viscosity, recently 
reviewed and parametrised by Huang {et al.}~\cite{HuangCryogenics2012}. First, we review how liquid $^{3}\mathrm{He}$ might be described as a lattice gas with a Hubbard model 
Hamiltonian.

Low temperature properties of liquid $^{3}\mathrm{He}$ can be described by Landau`s Fermi liquid theory. The effective mass of the quasi-particles [as deduced from the specific 
heat] is about 3 times the bare mass $m$ at 0 bar pressure and increases to 6 times at 33 bar, when the liquid becomes solid. The compressibility is also renormalised and 
decreases significantly with increasing pressure. This led Anderson and Brinkman to propose that $^3$He was an ``almost localised'' Fermi liquid. Thirty years ago, Vollhardt 
worked this idea out in detail, considering how these properties might be described by a lattice gas model with a Hubbard Hamiltonian.~\cite{VolhardtRMP1985} The system is at 
half filling with $U$ increasing with pressure, and the solidification transition (complete localisation of the fermions) then has some connection to the Mott transition. All of 
the calculations of Vollhardt were at the level of the Gutzwiller approximation (equivalent to Kotliar-Ruckenstein slave boson mean-field theory). A significant result from the 
theory is that it describes the weak pressure dependence and value of the Sommerfeld-Wilson ratio of the spin susceptibility to the specific heat [which is related to the Fermi 
liquid parameter $F_{0}^{a}$]. At ambient pressure $U$ was estimated to about 80 per cent of the critical value $U_c$ for the Mott transition. Vollhardt, Wolfle, and 
Anderson~\cite{VollhardtPRB87} also considered a more realistic situation where the system is not at half-filling. Then, the doping (band filling) is determined by the ratio of 
the molar volume of the liquid to the molar volume of the solid (which by definition corresponds to half filling). Finite temperatures extension to Volhardt theory was done by 
Seiler, Gros, Rice, Ueda, and Vollhardt.~\cite{Seiler} Later Georges and Laloux~\cite{Laloux} argued $^3$He is a Mott-Stoner liquid, i.e., one also needs to take into account the 
exchange interaction and proximity to a Stoner ferromagnetic instability. If this Mott-Hubbard picture is valid for $^3$He then one should also see a crossover from a Fermi liquid to a ``bad metal'' with increasing temperature. Specifically, above some ``coherence" temperature $T_{coh}$, the quasi-particle picture breaks down. For example, the specific heat 
per atom should increase linearly with temperature up to a value of order $k_B$  around $T_{coh}$, and then decrease with increasing temperature. Indeed one does see this 
crossover in experimental data (compare Figure 1 in Ref. \onlinecite{VollhardtPRL97}). 

We now consider what Hubbard model parameters are appropriate for $^3$He. The density at a pressure of 1 bar, $n\simeq N_{A}/(37\ \mathrm{cm}^{3})$ (where $N_{A}=6.023\times 
10^{23}$ is the Avagadro’s number) increases monotonically to $n \simeq N_{A}/(26\ \mathrm{cm}^{3})$ at 33 bar (near the solidification pressure) (see Table III in 
Ref.~\onlinecite{WheatleyRMP1975}).

The band mass $m_{b}$ can be written in terms of $E_{F} = \hbar^{2} k_{F}^{2}/(2m)$, the non-interacting Fermi energy, and the band energy $E_{b}$ as 
\begin{equation}
\frac{m}{m_{b}} = \frac{1}{2}(a k_{F})^2 \frac{E_{b}}{E_{F}}
= \frac{1}{2} (3 \pi^{2})^{2/3} \frac{E_{b}}{E_{F}}
\simeq 4.8 \frac{E_{b}}{E_{F}}
\end{equation}
where we have used the fact that $ n = 1/a^{3}= k_{F}^{3}/(3 \pi^{2})$.

There are several ways to estimate the band energy scale. If we have a Bethe lattice, then $E_{b} = W = E_{F}$, at half filling. Alternatively, we can compare the non-interacting 
density of states per spin  at the Fermi energy $N_{0}(0)$. This has the value of $3/(4 E_{F})$, $2/(\pi W)$, and $1/(\sqrt{\pi} t^{*})$ for the cases of a parabolic band (free 
fermions), Bethe lattice, and hypercubic lattice, respectively. Setting these equal gives $E_{b} \equiv W = 8 E_{F} /(3 \pi) \simeq 0.85 E_{F}$ and $E_{b} \equiv t^{*} = 4 E_{F} 
/(3 \sqrt{\pi}) \simeq 0.75 E_{F}$. Using the density at 1 bar and the non-interacting expression $E_{F} = \hbar^{2} k_{F}^{2}/(2m)$ we estimate $T_{F} \simeq 4.95$ K, and so 
$t^{*} \simeq 3.72 $ K.

In the following sections we compare some of our calculations of the shear viscosity with experimental data for $^{3}\mathrm{He}$.  Huang {et al.}~\cite{HuangCryogenics2012} 
showed that the shear viscosity of saturated liquid $^{3}\mathrm{He}$ from 3 mK to 0.1 K follows the Fermi liquid relation $\eta\propto 1/T^{2}$. Furthermore, they showed that 
the shear viscosity data in the range from 3 mK to near the critical point at 3.31 K, collected over the past 50 years from various experimental groups can be fitted to the 
empirical form :   
\begin{eqnarray}
\eta(T) = \frac{c_{1}}{T^{2}}+\frac{c_{2}}{T^{1.5}}+\frac{c_{3}}{T}+c_{4}
\label{eq:etaFit}
\end{eqnarray}
with $c_{1}=2.897\times 10^{-7}$ $\mathrm{Pa.s K}^{2}$, $c_{2}=-7.02\times10^{-7}$ $\mathrm{Pa.s K}^{1.5}$, $c_{3}=2.012\times10^{-6}$ $\mathrm{Pa.s K}$ and 
$c_{4}=1.323\times 10^{-6}$ $\mathrm{Pa.s}$. We note that at low temperatures Eq. (\ref{eq:etaFit}) has a Fermi liquid term. At high temperatures Eq. (\ref{eq:etaFit}) has the 
asymptotic value of $c_{4}$ which is comparable to $n \hbar \simeq 1.6 \times 10^{-6}$ $\mathrm{Pa.s}$ at 1 bar pressure. It should be pointed out that this is for data along the 
liquid-vapour curve and so the pressure gradually increases with temperature. However, as the critical pressure is about 100 kPa, much less than the melting pressure, this 
pressure dependence is not significant. The viscosity decreases by a factor of at most ten as the pressure increases from 1 kPa to 3 MPa [the melting pressure] for all 
temperatures below 1 K. Huang et al. fitted all the avaialable experimental data to an expansion in terms of Chebyshev polynomials and used this to plot the temperature 
dependence for pressures ranging from 1 kPa to 20 MPa (compare Figure 8 in Ref. \onlinecite{HuangCryogenics2012}). For pressures larger than about 500 kPa, the viscosity has a 
non-monotonic temperature dependence with a minimum around a temperature of 1 K.
\section{Results}
We consider the case of half-filling, $n=1$, i.e. each site on the average is occupied by one fermion. We study the shear viscosity and the entropy density as a function of 
correlation strength, $U$, and temperature, $T$.

\begin{figure}[!htbp]
\begin{center}
\includegraphics[scale=1.0,clip=]{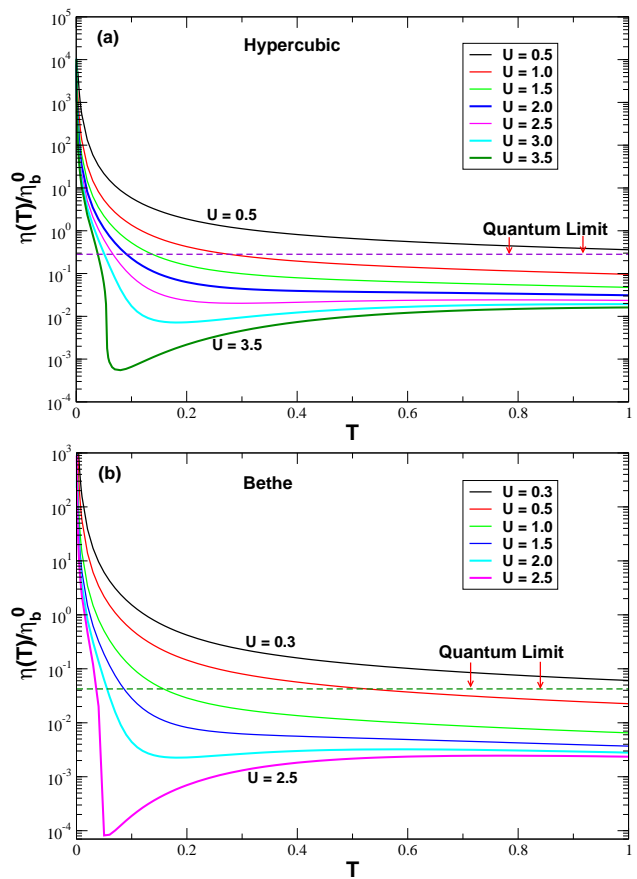}
\caption{(Color online) Shear viscosity, $\eta(T)/\eta_0^b$, as a function of temperature for a range of interaction strengths, $U$. Results are shown for both the hypercubic 
lattice [panel (a)] and the Bethe lattice [panel (b)]. Solid lines corresponds to DMFT based numerical results. $T$ and $U$ are measured in units of $W$ for the Bethe lattice and 
in units of $t^{*}$ for the hypercubic lattice case. Note that as the Mott insulator is approached the shear viscosity becomes extremely small. The dashed lines are the quantum 
limits given by Eq.~(\ref{Eq:EtaLatQuantumHyperCubic}) [hypercubic lattice] and Eq.~(\ref{Eq:EtaLatQuantumBethe}) [Bethe lattice].}
\label{fig:He3EtaPanel}
\end{center}
\end{figure}

In Fig.~\ref{fig:He3EtaPanel} (a) and Fig.~\ref{fig:He3EtaPanel}(b) we show the scaled shear viscosity $\eta(T)/\eta_{0}^{b}$ as a function of temperature for various interaction 
strengths $U$, for the hypercubic and the Bethe lattice, respectively. Similar results are obtained for both lattices.
\subsection{Quantum limits}
We consider violation of quantum limit of shear viscosity. In the weakly correlated hypercubic lattice system with $U=0.5$, the shear viscosity is always above quantum limit, 
$\eta_{q}^{lat} \simeq 0.28 \eta_{0}^{b}$ but as we increase the interaction strength, $U$, the shear viscosity smoothly goes below the quantum limit with increasing temperature, 
$T$. This corresponds to the fact that at low temperatures ($T \ll T_{coh}$) the quantum transport is due to coherent quasi-particle states but at high temperatures 
($T > T_{coh}$) the transport becomes incoherent in nature. This is the analogue of how in bad metals the resistivity smoothly increases above the Mott-Ioffe-Regel limit.
\subsection{Low temperature behaviour}
Fig.~\ref{fig:He3LogLogPlot} clearly shows that the shear viscosity follows Fermi liquid characteristic $1/T^{2}$ behaviour in the low temperature region ($T \ll T_{coh}$). 
The range of Fermi liquid behaviour decreases with increasing $U$. This is because the coherence scale [and Kondo temperature for the corresponding single impurity Anderson model]
decreases with increasing correlation strength $U$. The $1/T^{2}$ behaviour is similar to the low temperature behaviour of the electrical conductivity and the quantum transport in 
this region can be characterised by coherent quasi-particle states.    

Our calculated shear viscosity shows qualitative  behaviour consistent with experimental data for liquid $^3$He, using parameters estimated in the previous section. There is 
qualitative as well as quantitative agreement at low temperatures but not at higher temperatures. Interestingly, our calculated shear viscosity for $U=2$ for the hypercubic 
lattice nearly fits with the experimental results at low temperatures. Our calculation suggests $^{3}\mathrm{He}$ is a moderately correlated system with $U/U_{c}\sim 0.5$ 
[$U_{c}\sim 4.0$ for the hypercubic lattice] as against the suggestion of Volhardt~\cite{VolhardtRMP1985} that $^{3}\mathrm{He}$ is a nearly localized Fermi liquid with 
$U/U_{c}\sim 0.8$ at 1 bar pressure and $U/U_{c} \sim 0.9$ close to the melting pressure. It is important to mention that Gutzwiller based static mean field theory over estimates 
local correlation effects but the self-consistent treatment of dynamic correlation effects in DMFT renormalizes local correlation effects. 
\subsection{High temperature behaviour}
\begin{figure}[!htbp]
\begin{center}
\includegraphics[scale=0.34,clip=]{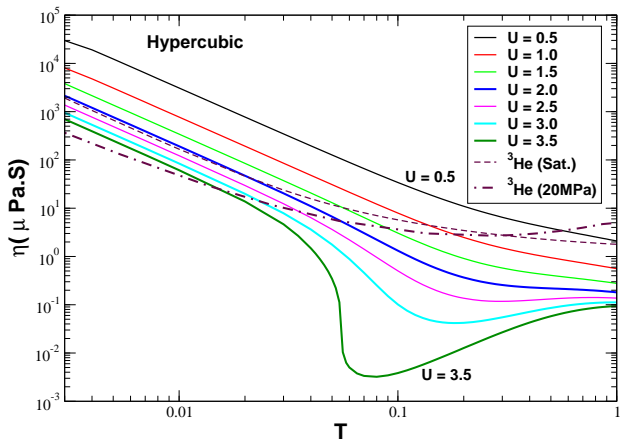}
\caption{(Color online) Shear viscosity, $\eta(T)$, on a log-log plot clearly shows  $1/T^{2}$ behaviour in the low temperature region, characteristic of a Fermi liquid. At high 
temperatures there is significant deviation from the Fermi liquid behaviour as the Mott transition is approached. The upper dashed line corresponds to experimental data for 
saturated liquid $^3$He parametrised by Eq. (\ref{eq:etaFit}). The lower dashed line is at a higher pressure. The results are shown for the hypercubic lattice case with the  
parameters $t^* = 3.72$ K and $m_b/m =3.6$. Both $T$ and $U$ are measured in units of $t^{*}$.}
\label{fig:He3LogLogPlot}
\end{center}
\end{figure}

In the high temperature region, $T\gg T_{coh}$, the shear viscosity shows significant deviation from the low temperature Fermi liquid behaviour as can be observed from 
Fig.~\ref{fig:He3EtaPanel} and Fig.~\ref{fig:He3LogLogPlot}. The quantum transport in this region is incoherent in nature. For the weakly and moderately correlated systems the 
deviation is smooth and monotonic but for strongly correlated systems for $U=2.5$ and above the deviation is much sharper and non-monotonic. This is due to the sharp crossover 
between the Fermi liquid fixed point and the local moment fixed point in the strongly correlated regime. A similar non-monotonic temperature dependence is seen in the electrical
resistivity from DMFT calculations and in organic charge charge transfer salts close to the Mott insulator.~\cite{MerinoPRB2000,LimelettePRL2003,PakhiraPRB2015}
\subsection{Entropy density}
\begin{figure}[!htbp]
\begin{center}
\includegraphics[scale=0.34,clip=]{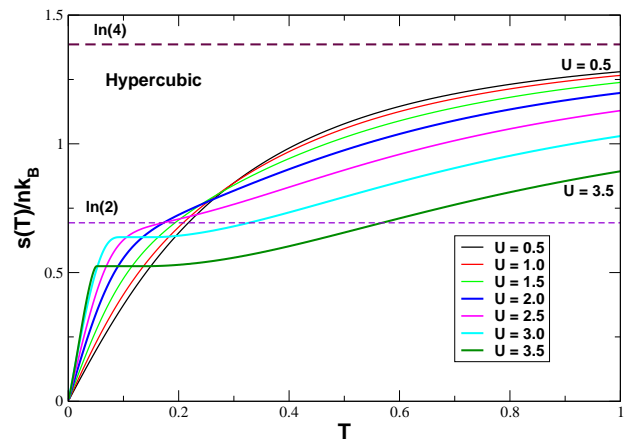}
\caption{(Color online) Entropy density, $s(T)$, (in units of $n k_{B}$) as a function of temperature, $T$, for various interaction strength, $U$. Below the coherence temperature, 
$T_{coh}$, the entropy is linear in temperature, characteristic of a Fermi liquid. The crucial point is that for $T >T_{coh}$, the entropy is of order $n k_B$. The kink like 
feature for $U=2.5$ and above corresponds to the formation of poorly screened local moment and its position is closely related to  $T_{coh}$. The calculation is for the hypercubic 
lattice and both $T$ and $U$ are measured in units of $t^{*}$.}
\label{fig:EntropyDensity}
\end{center}
\end{figure}

In Fig.~\ref{fig:EntropyDensity} we show the entropy density, $s(T)$, as a function of temperature for various interaction strengths. At high temperatures the entropy density 
approaches $\ln(4)$ which arises due to local charge and spin fluctuations. As the temperature decreases charge fluctuations freeze out and the model can be described by localised 
weakly interacting spin $1/2$’s with characteristic entropy density $\ln(2)$. Finally in the Fermi liquid state the local spin degrees of freedom are dynamically screened and the 
entropy density vanishes linearly in temperature. For weakly and moderately correlated electron system the entropy density smoothly crosses over $\ln(2)$. But for strongly 
correlated electron systems with $U = 2.5$ and above a kink like feature develops. This corresponds to formation of poorly screened local moment. The position of the kink in the
specific heat versus temperature curve is related to the coherence temperature, $T_{coh}$.\cite{ToschiPRL2009} For extremely correlated systems with $U=3.0$ and above the entropy 
density given by iterated perturbation theory (IPT) is under estimated. Consequently the specific heat in the coherent-incoherent crossover region becomes negative, which is 
unphysical. This is due to an incorrect total energy estimate in IPT which has been reported in earlier literature. (See for example, Figure 7 in Ref. \onlinecite{MoellerPRB1999})
\cite{Vucicevic}. In the unphysical temperature range we set the specific heat to zero and the calculated entropy density which is an integrated quantity will deviate by not more 
than 5\% from the actual value. Such a small error has little effect on whether the Kovtun-Son-Starinet (KSS) bound is violated.   
\subsection{Possible violation of the KSS bound}
Finally we consider the dimensionless scaled shear viscosity, $\eta(T)$, entropy density, $s(T)$, ratio
\begin{equation}
\zeta(T) \equiv \frac{\eta(T)}{s(T)}\frac{4 \pi k_B}{\hbar}\left(\frac{m_{b}}{m}\right)^{2}
\label{eqn:etakss}
\end{equation}
At the KSS bound $\zeta(T)=(\frac{m_{b}}{m})^{2}$ (for $d=3$). As stressed before this depends on the material properties $t^{*}$ and $a$ as well as the mass of the fermion, $m$.

\begin{figure}[!htbp]
\begin{center}
\includegraphics[scale=0.34,clip=]{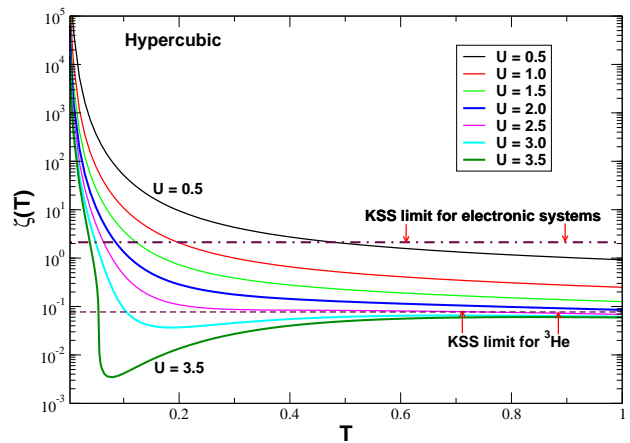}
\caption{(Color online) Dimensionless scaled ratio, $\zeta(T)$, of the shear viscosity and entropy density. [See Eq. (\ref{eqn:etakss}) for definition of $\zeta(T)$]. It is 
shown as a function of temperature, $T$, for various interaction strength, $U$. Dashed and dot-dashed lines corresponds to the Kovtun-Son-Starinets (KSS) limit for 
$^{3}\mathrm{He}$ and typical lattice electronic systems, respectively. For electronic systems there is strong violation of the quantum bound for $T \gg T_{coh}$. The calculation 
is for the hypercubic lattice case and both $T$ and $U$ are measured in units of $t^{*}$.}
\label{fig:EtaBySquantum}
\end{center}
\end{figure}

In Fig.~\ref{fig:EtaBySquantum} we show $\zeta(T)$ and compare to its value against the Kovtun-Son-Starinet (KSS) limit for parameters appropriate for $^{3}\mathrm{He}$ and 
typical lattice electronic systems like cuprates and organic superconductors.

For cuprates~\cite{PlatePRL2005} the hopping integral $t \simeq 0.18\mathrm{ eV}$, $a=3.9 \AA$ and for organic charge transfer salts\cite{PowellRepProgPhys2011} 
$t \simeq 0.05\mathrm{ eV}$, $a=8 \AA$. For these systems $d=2$ and hence $t^{*} = 2t\sqrt{d} \equiv 2\sqrt{2}t$. This will give $m/m_b \simeq 1.0 $ for cuprates and 
$m/m_b \simeq 1.2$ for organics, as compared to $m/m_{b} \simeq 3.6$ for liquid $^{3}$He. As a result the shear viscosity for these lattice electronic systems will be smaller by 
a factor of about 10 than for the charge neutral fermionic fluid $^{3}\mathrm{He}$. Straub and Harrison considered a simple model for the hopping integral for d-bands in 
transition metal.\cite{Harrison} This gives for $d$-sigma bands $m/m_{b} \simeq 2.8 (r_{d}/a)^{3}$, where $r_{d}$ is approximately the $d$-state radius and of the order of the 
radius transition metal atom, $\sim 1 \AA$. In principle then for a system with a large lattice constant, the band width can be very small and values of  $m/m_b$ even smaller 
than unity are possible. In an ultracold fermionic atom system in an optical lattice one could in principle then make $m/m_{b}$, and thus the viscosity, extremely small.

From Fig.~\ref{fig:EtaBySquantum} we can clearly see that for all $U < 3.0$ and for $^{3}$He parameters $\zeta(T)$ is above the KSS limit. For extremely 
correlated system $U=3.5$ there is strong violation of the limit in the crossover region but even for this system at high temperature the bound seems to be respected (within 
numerical error in calculation of entropy density, $s(T)$). Also, in the high temperature region the scaled ratio seems to approach some universal limit. 

For electronic lattice systems the limit is well respected in the coherent quasi-particle regime of transport but the limit is violated in the region $T > T_{coh}$. This is due 
to reduction of the shear viscosity by a factor of 10 compared to $^3$He parameters. The violation is as large as 1000\% for these systems, when they are close to the Mott 
transitions.  
\section{Experimental determination of $\eta$ in electronic systems}
Given our result that the KSS bound can be violated in a bad metal it is highly desirable that experimental measurements be performed on candidate strongly correlated electron 
materials such as organic charge transfer salts and cuprates.
Unfortunately, at present there is no direct measurement of the shear viscosity  for electronic systems. Recently an indirect estimate of $\eta/s$ was made from angle resolved 
photo-emission spectroscopy (ARPES) experiments in cuprates~\cite{RameauPRB2014} giving a value comparable to the KSS limit. However, it should be stressed that neither the 
viscosity nor the entropy were directly measured. Rather, the ARPES lineshape was used to estimate the quasi-particle lifetime and the state occupation. The viscosity was then 
estimated from the lifetime. The entropy was estimated from an expression in terms of the state occupations in a non-interacting fermion system. In the incoherent regime of 
transport this method will not be applicable. It is important to mention that our calculation showed that in the coherent quasi-particle regime of transport $\eta/s$ is always 
above the KSS bound and hence our result is consistent with these experimental results. 

A more direct way to measure the viscosity of the electron fluid in a metallic crystal is through the attenuation of sound, as first emphasized by Mason~\cite{Mason}. A more 
sophisticated and general theory was developed by Kahn and Allen~\cite{KhanPRB1987}. The connection between shear viscosity and ultrasound attenuation can be loosely motivated by 
Stokes law, given in Eq. (\ref{eqn-stokes}). In a metal, provided the wavelength of sound is much larger than the electronic mean free path, then one is in the hydrodynamic limit, and the attenuation is given by a similar expression to Stokes law (with appropriate indices for crystal axes), with $\rho$ the solid density, not that of the electron fluid. In 
a simple free electron model Eq. (\ref{eqn-fem}) shows that that the electronic viscosity is proportional to the scattering time, just like the electrical conductivity. Hence, 
the ultrasound attenuation should scale with the conductivity. Indirect evidence for this idea was found from the temperature dependence of ultrasound attenuation in 
aluminium~\cite{Lax}, including the predicted quadratic frequency dependence. In clean metals the attenuation (and viscosity) becomes very large at low temperatures, making it 
easier to measure. Also, for high frequency ultrasound, one can reach the ``quantum regime" where the mean free path becomes comparable to the sound wavelength. Pippard worked out 
a general theory describing the crossover from the hydrodynamic regime to this quantum regime~\cite{Pippard1955}. In bad metals could one experimentally measure the small 
electronic viscosity, of the order of $n \hbar$? First, the small mean free path, characteristic of bad metals, means one will always be in the hydrodynamic regime. However, the 
small viscosity means that the sound attenuation due to the electron fluid will be small and possibly dominated by other sources of attenuation such as crystal dislocations. A 
rough estimate for an electron viscosity of order of $n \hbar$, and a sound frequency of 1 GHz gives an attenuation of less than 0.1 cm$^{-1}$, of the order of typical 
sensitivity, such as in measurements for heavy fermion compounds~\cite{Batlogg1986}.

Resonant Ultrasound Spectroscopy (RUS) \cite{Migliori,LeisureJPhysCondMat1997,MigloriRSI2005} has been used to make measurements on strongly correlated electron 
systems~\cite{MiglioriPhysicaB1994,CarpenterPRB2010,ShekhterNature2013}. The spectrum is determined by the resonant elastic modes of the sample; they are determined by the sample 
shape and orientation, elastic constants, and dissipation. RUS allows determination of the elastic constant tensor $C_{ij}$ from measurements on small samples ($<1$ mm$^3$ volume).
In the regime where the attenuation of the ultrasound is dominated by coupling to the electrons, rather than fluctuations associated with phase transitions, the viscosity could 
be determined from the damping [frequency width, $\Delta \omega$] of the resonances. We estimate  $\Delta \omega / \omega \sim \eta \omega/C_{ij}$ \cite{Bhatia} and so for the 
MHz frequencies typically used in RUS, the damping associated with a viscosity of order $n \hbar$ requires an oscillator Q factor of order $10^{10}$ and so is unlikely to be 
observable.

Recently, several new approaches have been suggested to experimentally measure the viscosity of the electron fluid in a metallic crystal. Forcella, Zaanen, Valentinis, and van 
Der Marel~\cite{ForcellaPRB2014} considered electromagnetic properties of viscous charged fluids, finding signatures due to the viscosity such as negative refraction, a frequency 
dependent peak in the reflection coefficient, and a strong frequency dependence of the phase. However, they note that these effects may be difficult to observe for viscosities of 
the order of $n \hbar$. Tomadin, Vignale, and Polini~\cite{TomadinPRL2014} considered a two-dimensional electron fluid in a Corbino disk device in the presence of an oscillating 
magnetic flux. They showed that the viscosity could be determined from the dc potential difference that arises between the inner and the outer edge of the disk.  In particular, 
for viscosities of the order of $n \hbar$ the potential difference varied significantly  oscillation frequencies in the MHz range. Levitov and Falkovich~\cite{Levitov} recently
considered the flow of an electron fluid in a micrometer scale channel in the hydrodynamic regime, where the electron-electron collision rate is much larger than the momentum 
relaxation rate. They found that when the viscosity to resistance ratio is sufficiently large viscous flow occurs producing vorticity and a negative nonlocal voltage. Spatially 
resolved measurements of the voltage allow determination of the viscosity. Torre, Tomadin, Geim, and Polini~\cite{Torre} considered the electron liquid in graphene in the 
hydrodynamic regime and showed that the shear viscosity could be determined from measurements of non-local resistances in multi-terminal Hall bar devices. Although these proposals are promising for the two-dimensional electron fluids in graphene and semiconductor heterostructures fabrication of the relevant micron-scale devices may be particularly 
challenging for bad metals such as cuprates and organic charge transfer salts.
\section{Conclusions}
We have studied the shear viscosity, the entropy density and their ratio for a single band Hubbard model using single site Dynamical Mean Field Theory. Similar results were 
obtained for the density of states associated with both hypercubic and Bethe lattices. We compared our results for the temperature dependence of the shear viscosity to 
experimental results for liquid $^{3}\mathrm{He}$. The calculated shear viscosity shows qualitative as well as quantitative behaviour consistent with experimental results. At low 
temperatures the shear viscosity is proportional to $1/T^{2}$ corresponding to coherent quasi-particle based transport in the Fermi liquid state. At high temperatures the shear 
viscosity shows significant deviation from Fermi liquid state behavior. This corresponds to crossover from coherent quasi-particle based transport to incoherent transport (the 
``bad metal''). With increasing interaction strength $U$ the shear viscosity becomes less than conjectured quantum limits of shear viscosity, of the order of $n \hbar$.  Finally, 
we considered the scaled dimensionless ratio between shear viscosity and entropy density. This ratio in the Hubbard model depends on the energy scale $t^{*}$, length scale $a$, 
and the free fermion mass $m$. This is in contrast to the universal limit $\frac{\hbar}{4\pi k_{B}}$ predicted by Kovtun, Son, and Starinets using the AdS/CFT correspondence in a 
conformally symmetric field theory model. For $^{3}\mathrm{He}$ parameters the ratio is  above the universal bound but for parameters appropriate for electronic lattice systems, 
such as cuprate and organic metals, this bound is found to be strongly violated, in the bad metal regime near the Mott metal-insulator transition.  
We hope that our results will stimulate experimental measurements of the shear viscosity in bad metals.
\section*{Acknowledgments} 
We would like to acknowledge useful discussions with R. Adhikari,  J. Analytis, A. Andreev, G. Baskaran, A. Bulgac, V. Dobrosavljevic, S. Hartnoll, S. Kalyan Rama, M. Laad, 
R. Mann, B. Sathiapalan,  T. Sch\"afer, B. Spivak, D. Tanaskovic, D. Vollhardt, J. Vucicevic, and G. Wlazlowski. We thank Y. Huang for providing data from Reference 
\onlinecite{HuangCryogenics2012}. We thank M. Kollar for drawing our attention to analytical derivation of Eq.~(\ref{eq:ThetaHyp}). This work was supported by a Discovery Project 
grant from the Australian Research Council.
\section*{APPENDIX: Transport density of states for shear viscosity}
In the limit $d\rightarrow\infty$ the viscosity will involve the following transport function
\begin{eqnarray}
\Theta_{xy}(\epsilon)= m^{2} \sum_{\mathbf{k}} v^{2}_{x}v^{2}_{y}\delta(\epsilon-\epsilon_{\mathbf{k}}).
\end{eqnarray}
For a $d$-dimensional hypercubic lattice
\begin{eqnarray}
\Theta_{xy}(\epsilon)=\frac{m^{2}(2t)^{4}a^{4}}{\hbar^{4}}\sum_{\mathbf{k}} \sin^{2}(k_{x})\sin^{2}(k_{y})\delta(\epsilon-\epsilon_{\mathbf{k}}).
\end{eqnarray} 
To evaluate this we first we define the Fourier transform
\begin{eqnarray}
Y(\omega) &=& \int\limits_{-\infty}^{+\infty} \Theta_{xy}(\epsilon) e^{-i\omega\epsilon} d\epsilon \nonumber \\
          &=& \gamma^{2}(2t)^{4}\sum_{\mathbf{k}} \sin^{2}(k_{1}) \sin^{2}(k_{2}) \prod\limits_{\alpha=1}^{d} e^{i2t\omega\cos(k_{\alpha})}\nonumber \\
          &=& \gamma^{2}(2t)^{4} J_{0}^{d-2}(2t\omega)\left[\frac{J_{1}(2t\omega)}{2t\omega}\right]^{2}\nonumber \\
          &=& \frac{(2t\gamma)^{2}}{\omega^{2}} J_{0}^{d-2}(2t\omega)\left[J_{1}(2t\omega)\right]^{2}
\end{eqnarray}
where $\gamma\equiv\frac{ma^{2}}{\hbar^{2}}$.

Using relations for Bessel functions
\begin{eqnarray}
&&\left[J_{0}(2t\omega)\right]^{d-2}\left[J_{1}(2t\omega)\right]^{2} = \frac{1}{(2t)^{2}d(d-1)}\frac{d^{2}J_{0}^{d}(2t\omega)}{d\omega^{2}}\nonumber \\
&&+\frac{1}{d-1} J_{0}^{d}(2t\omega)-\frac{1}{2t(d-1)}\frac{1}{\omega}\left[J_{0}(2t\omega)\right]^{d-1}J_{1}(2t\omega) 
\end{eqnarray} 
we can rewrite $Y(\omega)$ as
\begin{eqnarray}
Y(\omega)&=&\frac{\gamma^{2}}{d(d-1)}\frac{1}{\omega^{2}} \frac{d^{2}J_{0}(2t\omega)}{d\omega^{2}}+\frac{(2t\gamma)^{2}}{(d-1)}\frac{1}{\omega^{2}} J_{0}^{d}(2t\omega)\nonumber \\
         &-&\frac{2t\gamma^{2}}{(d-1)}\frac{1}{\omega^{3}}\left[J_{0}(2t\omega)\right]^{d-1}J_{1}(2t\omega)\nonumber\\
         &\equiv& Y_{1}(\omega)+Y_{2}(\omega)+Y_{3}(\omega).
\end{eqnarray}
We can Fourier transform back to calculate $\Theta_{xy}(\epsilon)$ as
\begin{eqnarray}
\Theta_{xy}(\epsilon) &=& \frac{1}{2\pi}\int\limits_{-\infty}^{+\infty} Y(\omega) e^{i\omega\epsilon} d\omega.
\end{eqnarray}
Using the convolution theorem we can easily show that each term of $Y(\omega)$ has the following form
\begin{eqnarray}
 \Theta_{xy}^{(\alpha)}(\epsilon)&=& \int\limits_{-\infty}^{+\infty} F_{\alpha}(\epsilon-z)G_{\alpha}(z) dz, \hspace*{0.25cm} \alpha=1,\cdots,3
\end{eqnarray}
where,
\begin{eqnarray}
F_{\alpha}(\epsilon)=\frac{1}{2\pi}\int_{-\infty}^{+\infty}\tilde{F}_{\alpha}(\omega)e^{i\omega\epsilon}d\omega\nonumber \\
G_{\alpha}(\epsilon)=\frac{1}{2\pi}\int_{-\infty}^{+\infty}\tilde{G}_{\alpha}(\omega)e^{i\omega\epsilon}d\omega,
\end{eqnarray}
and $Y_{\alpha}(\omega)=\tilde{F}_{\alpha}(\omega)\tilde{G}_{\alpha}(\omega)$. 

For the first term 
\begin{eqnarray}
F_{\alpha}(\epsilon) &=& \frac{1}{2\pi}\int\limits_{-\infty}^{+\infty} \frac{1}{\omega^{2}} e^{i\omega\epsilon}d\omega\nonumber \\
                     &=& \frac{1}{2\pi}\cdot \pi i \cdot i\epsilon \; sgn(\epsilon) = -\frac{\epsilon}{2} sgn(\epsilon)
\end{eqnarray}
\begin{eqnarray}
G_{\alpha}(\epsilon) &=& \frac{1}{2\pi}\int\limits_{-\infty}^{+\infty} \frac{d^{2}[J_{0}(2t\omega)]^{d}}{d\omega^{2}} e^{i\omega\epsilon}d\omega \nonumber \\
                     &=& -\epsilon^{2} N_{0}(\epsilon)
\end{eqnarray}
where we have used
\begin{eqnarray}
& & \left . \frac{d[J_{0}(2t\omega)]^{d}}{d\omega}e^{i\omega\epsilon}\right|_{-\infty}^{+\infty} = 0 \nonumber \\
& & \left . [J_{0}(2t\omega)]^{d}e^{i\omega\epsilon}\right|_{-\infty}^{+\infty} = 0.
\end{eqnarray}
Finally we get
\begin{eqnarray}
\Theta_{xy}^{(1)} (\epsilon) &=& \frac{\gamma^{2}\epsilon}{d(d-1)} \int\limits_{-\infty}^{\epsilon} z^{2} N_{0}(z) dz - \frac{(2t\gamma)^{2}\epsilon}{4(d-1)}\nonumber \\ 
                             &-& \frac{\gamma^{2}}{d(d-1)} \int\limits_{-\infty}^{\epsilon} z^{3} N_{0}(z) dz,
\end{eqnarray}
where we have used $\int_{-\infty}^{+\infty} z^{2} N_{0}(z) dz=\sum_{\mathbf{k}} \epsilon_{\mathbf{k}}^{2}=2t^{2}d$ and $\int_{-\infty}^{+\infty} z^{3} N_{0}(z) dz=0$.

A similar exercise for the second term will give
\begin{eqnarray}
\Theta_{xy}^{(2)}(\epsilon) &=& -\frac{(2t\gamma)^{2}\epsilon}{(d-1)} \int\limits_{-\infty}^{\epsilon} N_{0}(z) dz + \frac{(2t\gamma)^{2}\epsilon}{2(d-1)}\nonumber \\
                            &+& \frac{(2t\gamma)^{2}}{(d-1)}\int\limits_{-\infty}^{\epsilon} z N_{0}(z) dz.
\end{eqnarray}

For the third term we have
\begin{eqnarray}
F_{3}(\epsilon) &=& \frac{1}{2\pi} \int\limits_{-\infty}^{+\infty} \frac{1}{\omega^{3}} e^{i\omega\epsilon}d\omega\nonumber \\
          &=& \lim\limits_{\omega\rightarrow 0}\frac{1}{2\pi}\cdot \pi i \cdot \frac{1}{2!} \frac{d^{2}}{d\omega^{2}}\left[e^{i\omega\epsilon}\right]sgn(\epsilon)\nonumber \\
          &=& - i \frac{\epsilon^{2}}{4} sgn(\epsilon)
\end{eqnarray}
and
\begin{eqnarray}
G_{3}(\epsilon) &=& -\frac{1}{2\pi} \frac{2t\gamma^{2}}{(d-1)}\int\limits_{-\infty}^{+\infty} \left[J_{0}(2t\omega)\right]^{d-1} J_{1}(2t\omega) d\omega\nonumber \\
                &=& \frac{1}{2\pi}\frac{\gamma^{2}}{d(d-1)}\int\limits_{-\infty}^{+\infty} \frac{d[J_{0}(2t\omega)]^{d}}{d\omega}d\omega \nonumber \\
                &=& -i\frac{\gamma^{2}}{d(d-1)}\epsilon N_{0}(\epsilon). 
\end{eqnarray}
Finally, we have
\begin{eqnarray}
\Theta_{xy}^{(3)}(\epsilon)&=& - \frac{\gamma^{2}\epsilon^{2}}{2d(d-1)}\int\limits_{-\infty}^{\epsilon} z N_{0}(z) dz -\frac{(2t\gamma)^{2}\epsilon}{4(d-1)}\nonumber \\ 
                           & &\hspace*{-2cm}+\frac{\gamma^{2}\epsilon}{d(d-1)}\int\limits_{-\infty}^{\epsilon} z^{2}N_{0}(z) dz 
                                -\frac{\gamma^{2}}{2d(d-1)}\int\limits_{-\infty}^{\epsilon} z^{3}N_{0}(z) dz.
\end{eqnarray}
Collecting and rearranging all the terms we obtain Eq. (\ref{eqn:thetaxy2}).


\begin{thebibliography}{10}

\bibitem{Bhatia}
A.B. Bhatia, {\it Ultrasonic absorption: An introduction to the Theory of Sound Absorption
and Dispersion in Gases, Liquids and Solids}
(Oxford, 1967), page 54.

\bibitem{HuangCryogenics2012}
Y.~Huang, Q.~Yu, Q.~Chen, and R.~Wang,
\newblock Cryogenics {\bf 52}, 538 (2012).

\bibitem{SteinbergPR1958}
M.~S. Steinberg,
\newblock Phys. Rev. {\bf 109}, 1486 (1958).

\bibitem{EyringJCP1936}
H.~Eyring,
\newblock J. Chem. Phys. {\bf 4}, 236 (1936).

\bibitem{MerinoPRB2000}
J.~Merino and R.~H. McKenzie,
\newblock Phys. Rev. B {\bf 61}, 7996 (2000).

\bibitem{HusseyPhilMag2004}
N.~E. Hussey, K.~Takenaka, and H.~Takagi,
\newblock Phil. Mag. {\bf 84}, 2847 (2004).

\bibitem{Calandra}
O.~Gunnarsson, M.~Calandra, and J.~E. Han,
\newblock Rev. Mod. Phys. {\bf 75}, 1085 (2003).

\bibitem{SachdevJPcond2009}
S.~Sachdev and M.~M\"{u}ller,
\newblock J. Phys : Condens. Matter {\bf 21}, 164216 (2009).

\bibitem{FaulknerScience2010}
T.~Faulkner, N.~Iqbal, H.~Liu, J.~McGreevy, and D.~Vegh,
\newblock Science {\bf 329}, 1043 (2010).

\bibitem{LiuPT2012}
H.~Liu,
\newblock Phys. Today {\bf 65}, 68 (2012).

\bibitem{HartnollNatPhys2014}
S.~A. Hartnoll,
\newblock Nat. Phys. {\bf 11}, 54 (2014).

\bibitem{Zaanen}
R.~A. Davison, K.~Schalm, and J.~Zaanen,
\newblock Phys. Rev. B {\bf 89}, 245116 (2014).

\bibitem{Maldacena1998}
J.~M. Maldacena,
\newblock Adv. Theor. Math. Phys. {\bf 2}, 231 (1998).

\bibitem{KovtunPRL2005}
P.~K. Kovtun, D.~T. Son, and A.~O. Starinets,
\newblock Phys. Rev. Lett. {\bf 94}, 111601 (2005).

\bibitem{Shuryak2004}
E.~Shuryak,
\newblock Prog. Part. Nucl. Phys. {\bf 53}, 273 (2004).

\bibitem{CaoScience2011}
C.~Cao, E.~Elliott, J.~Joseph, H.~Wu, J.~Petricka, T.~Sch\"{a}fer, and J.~E.
  Thomas,
\newblock Science {\bf 331}, 58 (2011).

\bibitem{MullerPRL2009}
M.~M\"uller, J.~Schmalian, and L.~Fritz,
\newblock Phys. Rev. Lett. {\bf 103}, 025301 (2009).

\bibitem{Chafin}
C.~Chafin and T.~Sch\"afer,
\newblock Phys. Rev. A {\bf 87}, 023629 (2013).

\bibitem{Bulgac}
G.~Wlaz\l{}owski, P.~Magierski, A.~Bulgac, and K.~J. Roche,
\newblock Phys. Rev. A {\bf 88}, 013639 (2013).


\bibitem{ElliottPRL2014}
E. Elliott, J.A. Joseph, and J.E. Thomas, 
Phys. Rev. Lett. {\bf 113}, 020406 (2014).

\bibitem{Wlazlowski}
G. Wlazlowski, W. Quan, and A. Bulgac,
arXiv:1504.02560.

\bibitem{Myers}
T. D. Cohen, 
Phys. Rev. Lett. {\bf 99}, 021602 (2007); 
A. Sinha and R.C. Myers,
Nuclear Physics A {\bf 830}, 295c (2009);
S. Cremonini,
Mod. Phys. Lett. B {\bf 25}, 1867 (2011).

\bibitem{PakhiraPRB2015}
N.~Pakhira and R.~H. McKenzie,
\newblock Phys. Rev. B {\bf 91}, 075124 (2015).

\bibitem{SchaferAnnRevNuclPart2014}
T.~Sch\"{a}fer,
\newblock Ann. Rev. Nucl. Part. Sci. {\bf 64}, 125 (2014).

\bibitem{BruunPRA2007}
G.~M. Bruun and H.~Smith,
\newblock Phys. Rev. A {\bf 75}, 043612 (2007).

\bibitem{TaylorPRA2010}
E.~Taylor and M.~Randeria,
\newblock Phys. Rev. A {\bf 81}, 053610 (2010).

\bibitem{ashcroft1976solid}
N.~Ashcroft and N.~Mermin,
\newblock {\em Solid State Physics} (Saunders College, 1976).

\bibitem{KhanPRB1984}
F.~S. Khan and P.~B. Allen,
\newblock Phys. Rev. B {\bf 29}, 3341 (1984).

\bibitem{KhanPRB1987}
F.~S. Khan and P.~B. Allen,
\newblock Phys. Rev. B {\bf 35}, 1002 (1987).

\bibitem{Rah}
K. Rah and B.C. Eu,
Phys. Rev. E {\bf 60}, 4105 (1999).

\bibitem{Enss}
T. Enss, R. Haussmann, and W. Zwerger,
Annals Phys. {\bf 326}, 770 (2011).

\bibitem{MetznerPRL1989}
W.~Metzner and D.~Vollhardt,
\newblock Phys. Rev. Lett. {\bf 62}, 324 (1989).

\bibitem{GeorgesRMP1995}
A.~Georges, G.~Kotliar, W.~Krauth, and M.~J. Rozenberg,
\newblock Rev. Mod. Phys. {\bf 68}, 13 (1996).

\bibitem{LimelettePRL2003}
P.~Limelette, P.~Wzietek, S.~Florens, A.~Georges, T.~A. Costi, C.~Pasquier,
  D.~J\'erome, C.~M\'ezi\`ere, and P.~Batail,
\newblock Phys. Rev. Lett. {\bf 91}, 016401 (2003).

\bibitem{MerinoPRL2008}
J.~Merino, M.~Dumm, N.~Drichko, M.~Dressel, and R.~H. McKenzie,
\newblock Phys. Rev. Lett. {\bf 100}, 086404 (2008).

\bibitem{PowellRepProgPhys2011}
B.~J. Powell and R.~H. McKenzie,
\newblock Rep. Prog. Phys. {\bf 74}, 056501 (2011).

\bibitem{KotliarRMP2006}
G.~Kotliar, S.~Y. Savrasov, K.~Haule, V.~S. Oudovenko, O.~Parcollet, and C.~A.
  Marianetti,
\newblock Rev. Mod. Phys. {\bf 78}, 865 (2006).

\bibitem{ZhangPRL1993}
X.~Y. Zhang, M.~J. Rozenberg, and G.~Kotliar,
\newblock Phys. Rev. Lett. {\bf 70}, 1666 (1993).

\bibitem{KajueterPRL1996}
H.~Kajueter and G.~Kotliar,
\newblock Phys. Rev. Lett. {\bf 77}, 131 (1996).

\bibitem{BullaPRL1999}
R.~Bulla,
\newblock Phys. Rev. Lett. {\bf 83}, 136 (1999).

\bibitem{TerletskaPRL2011}
H.~Terletska, J.~Vu\ifmmode \check{c}\else \v{c}\fi{}i\ifmmode \check{c}\else
  \v{c}\fi{}evi\ifmmode~\acute{c}\else \'{c}\fi{},
  D.~Tanaskovi\ifmmode~\acute{c}\else \'{c}\fi{}, and
  V.~Dobrosavljevi\ifmmode~\acute{c}\else \'{c}\fi{},
\newblock Phys. Rev. Lett. {\bf 107}, 026401 (2011).

\bibitem{SachdevPRL}
S. Sachdev,
 Phys. Rev. Lett. {\bf 105}, 151602 (2010).

\bibitem{KhuranaPRL1990}
A.~Khurana,
\newblock Phys. Rev. Lett. {\bf 64}, 1990 (1990).

\bibitem{ArsenaultPRB2013}
L.-F. Arsenault and A.-M.~S. Tremblay,
\newblock Phys. Rev. B {\bf 88}, 205109 (2013).

\bibitem{Aarts}
G. Aarts and J.M. Martinez Resco,
J. High Energy Phys. {\bf 4}, 53 (2002).

\bibitem{Novikov}
D.S. Novikov, arXiv:cond-mat/0603184.

\bibitem{Lin}
N.~Lin, E.~Gull, and A.~J. Millis,
 Phys. Rev. B {\bf 80}, 161105 (2009).

\bibitem{BergeronPRB2011}
D.~Bergeron, V.~Hankevych, B.~Kyung, and A.-M.~S. Tremblay,
Phys. Rev. B {\bf 84}, 085128 (2011).

\bibitem{Economou2010Green}
E.~Economou,
\newblock {\em Green's Functions in Quantum Physics} (Springer Berlin
  Heidelberg, 2010).

\bibitem{Georges1993}
A.~Georges and W.~Krauth,
\newblock Phys. Rev. B {\bf 48}, 7167 (1993).

\bibitem{PruschkePRB1993}
T.~Pruschke, D.~L. Cox, and M.~Jarrell,
\newblock Phys. Rev. B {\bf 47}, 3553 (1993).

\bibitem{MoellerPRB1999}
G.~Moeller, V.~Dobrosavljevi{\'c}, and A.~E. Ruckenstein,
\newblock Phys. Rev. B {\bf 59}, 6846 (1999).

\bibitem{VolhardtRMP1985}
D.~Vollhardt,
\newblock Rev. Mod. Phys. {\bf 56}, 99 (1984).

\bibitem{VollhardtPRB87}
D.~Vollhardt, P.~W\"olfle, and P.~W. Anderson,
\newblock Phys. Rev. B {\bf 35}, 6703 (1987).

\bibitem{Seiler}
K.~Seiler, C.~Gros, T.~Rice, K.~Ueda, and D.~Vollhardt,
\newblock J. Low Temp. Phys. {\bf 64}, 195 (1986).

\bibitem{Laloux}
A.~Georges and L.~Laloux,
\newblock Mod. Phys. Lett. B {\bf 11}, 913 (1997).

\bibitem{VollhardtPRL97}
D.~Vollhardt,
\newblock Phys. Rev. Lett. {\bf 78}, 1307 (1997).

\bibitem{WheatleyRMP1975}
J.~C. Wheatley,
\newblock Rev. Mod. Phys. {\bf 47}, 415 (1975).

\bibitem{ToschiPRL2009}
A.~Toschi, M.~Capone, C.~Castellani, and K.~Held,
\newblock Phys. Rev. Lett. {\bf 102}, 076402 (2009).

\bibitem{Vucicevic}
J.~Vucicevic,
\newblock unpublished  (2015).

\bibitem{PlatePRL2005}
M.~Plat\'e {\em et~al.},
\newblock Phys. Rev. Lett. {\bf 95}, 077001 (2005).

\bibitem{Harrison}
G.~K. Straub and W.~A. Harrison,
\newblock Phys. Rev. B {\bf 31}, 7668 (1985).

\bibitem{RameauPRB2014}
J.~D. Rameau, T.~J. Reber, H.-B. Yang, S.~Akhanjee, G.~D. Gu, P.~D. Johnson,
  and S.~Campbell,
\newblock Phys. Rev. B {\bf 90}, 134509 (2014).

\bibitem{Mason}
W.~P. Mason,
\newblock Phys. Rev. {\bf 97}, 557 (1955).

\bibitem{Lax}
E.~Lax,
\newblock Phys. Rev. {\bf 115}, 1591 (1959).

\bibitem{Pippard1955}
A.~Pippard,
\newblock Phil. Mag. {\bf 46}, 1104 (1955).

\bibitem{Batlogg1986}
B.~Batlogg, D.~Bishop, B.~Golding, E.~Bucher, J.~Hufnagl, Z.~Fisk, J.~Smith,
  and H.~Ott,
\newblock Phys. Rev. B {\bf 33}, 5906 (1986).

\bibitem{Migliori}
A. Migliori and J.L. Sarrao,
{\it Resonant Ultrasound Spectroscopy: Applications to Physics, Materials Measurements, and Nondestructive Evaluation},
(Wiley-VCH, 1997).

\bibitem{LeisureJPhysCondMat1997}
R.G. Leisure and F.A. Willis,
J. Phys.: Condens. Matter {\bf 9}, 6001 (1997).

\bibitem{MigloriRSI2005} 
A. Migliori and J.D. Maynard,
Rev. Sci. Instrum. {\bf 76}, 121301 (2005).

\bibitem{MiglioriPhysicaB1994}
J.L. Sarrao, D. Mandrus, A. Migliori, Z. Fisk, and E. Bucher,
Physica B {\bf  199 - 200},   478 (1994).

\bibitem{CarpenterPRB2010}
M.A. Carpenter, C.J. Howard, R.E.A. McKnight, A. Migliori, J.B. Betts, and V.R. Fanelli,
Phys. Rev. B {\bf 82}, 134123 (2010).

\bibitem{ShekhterNature2013}
A. Shekhter,	B.J. Ramshaw,	R. Liang, W.N. Hardy,	D.A. Bonn, F.F. Balakirev, 
R.D. McDonald,	J.B. Betts,	S.C. Riggs, and A. Migliori,
Nature {\bf 498}, 75 (2013).

\bibitem{ForcellaPRB2014}
D. Forcella, J. Zaanen, D. Valentinis, and D. van Der Marel,
Phys. Rev. B {\bf 90}, 035143 (2014).

\bibitem{TomadinPRL2014}
A. Tomadin, G.  Vignale, and M. Polini,
Phys. Rev. Lett. {\bf 113}, 235901 (2014).

\bibitem{Levitov}
L. Levitov and  G. Falkovich,
arXiv:1508.00836.

\bibitem{Torre}
I. Torre, A. Tomadin, A.K. Geim, and M. Polini,
arXiv:1508.00363.

\end{thebibliography}
\end{document}